\documentclass[12pt]{article}
\usepackage{amssymb,amsmath,epsfig}

\begin{document}

\title{\bf Evolution of Axially Symmetric Anisotropic Sources in $f(R,T)$ Gravity}

\author{M. Zubair
\thanks{mzubairkk@gmail.com; drmzubair@ciitlahore.edu.pk} ${}^{(a)}$,
Ifra Noureen \thanks{ifra.noureen@gmail.com} ${}^{(b)}$ \\\
${}^{(a)}$ Department of Mathematics,\\ COMSATS Institute of Information
Technology, Lahore, Pakistan.\\ ${}^{(b)}$ Department of
Mathematics,\\University of Management and Technology, Lahore, Pakistan.}
\date{}
\maketitle

\begin{abstract}
We discuss the dynamical analysis in $f(R,T)$ gravity (where $R$ is Ricci
scalar and $T$ is trace of energy momentum tensor) for gravitating sources
carrying axial symmetry. The self gravitating system is taken to be
anisotropic and line element describes axially symmetric geometry avoiding
rotation about symmetry axis and meridional motions (zero vorticity case).
The modified field equations for axial symmetry in $f(R,T)$ theory are
formulated, together with the dynamical equations. Linearly perturbed
dynamical equations lead to the evolution equation carrying adiabatic index
$\Gamma$ that defines impact of non-minimal matter to geometry coupling on
range of instability for Newtonian (N) and post-Newtonian (pN)
approximations.
\end{abstract}


\section{Introduction}

Recent developments in astrophysics and structure formation theories reveals
that gravitating sources might deviate from most commonly studied spherical
symmetry. Such deviations in realistic scenarios appear incidently, giving
rise to the worth of non-spherical symmetries in gravitating objects. Herein,
we intend to look into the implications of restricted class of axially
symmetric sources (avoiding reflection and rotation) on gravitational
evolution in context of $f(R,T)$ theory of gravity. Consideration of dynamic
sources together with the angular momentum is a cumbersome task, however,
observational data suggests that the lack of spherical symmetry prevails the
more practical and worthwhile situations. A viable $f(R,T)$ model
($\frac{df}{dr}\geq0, \frac{d^{2}f}{dr^{2}}\geq0$) with locally anisotropic
matter distribution has been taken into account for the dynamical analysis.

The evolution of gravitating sources has been studied with a great deal of
interest in recent past. Stars tend to collapse when outward drawn pressure
decreases because of continuous fuel consumption, leading to imbalance in
outward forces and inward acting gravitational pull. In such situation,
gravitational force becomes the only governing force, massive stars burn
nuclear fuel more rapidly and so more unstable as compare to the stars with
relatively less mass. There are many other factors other than mass of the
gravitating source that implicate intense modifications in range of
stability/instability such as isotropy, anisotropy, shear, dissipation and
radiation. Chanderashekar \cite{1} shared valuable explorations to set
instability range for spherically symmetric gravitating source in the form of
adiabatic index $\Gamma$ comprising pressure to density ratio with time
transition.

Hillebrandt and Steinmetz \cite{2} presented the instability criterion for
anisotropic matter configuration of gravitating objects. Herrera et al.
\cite{3}-\cite{7} contributed majorally in establishment of instability range
of general relativistic fluids for different cases (isotropic, anisotropic,
dissipative collapse etc), remarked that pressure anisotropy largely
participates in setting dynamical instability. Moreover, they also worked out
the imprints of axially and reflection symmetric static and dynamic sources
by a general framework and some analytic models \cite{8}. Axially symmetric
shearing geodesic and shearfree dissipative fluids are discussed in
\cite{9,10}, where shearing geodesic case represents the zero radiation
production.

General Relativity (GR) is a self-consistent theory, it is adequate for the
explanation of many gravitational phenomenons up to cosmological scales. The
scheme of GR appears to disagree with progressing observational data such as
large scale structures ranging from galaxies to galaxy clustering, IA-type
Supernovae, cosmic microwave background \cite{11}-\cite{14} etc.
Alternatively, it can be said that GR is not the only definite gravitational
theory that is suitable for all scales. Many attempts have been made to
validate gravitational theories on large scales and coup with the cosmic
acceleration \cite{15}-\cite{21}, by introducing modified theories of gravity
\cite{22}-\cite{27} for e.g. $f(R)$, $f(G)$, Brans-Dicke theory, $f(R,T)$ and
so on.

Since the introduction of $f(R,T)$ theory in 2011 \cite{28}, people
\cite{29}-\cite{31} worked on energy conditions along with its cosmological
and thermodynamic implications. The $f(R,T)$ theory represents generalization
of $f(R)$ theory carrying non-minimal matter to geometry coupling. Extensive
work has been done on instability range of spherically symmetric stars in GR
as well as in modified theories of gravity. Literature on dynamical analysis
of axially symmetric sources can be witnessed in GR. However, being a heavier
task to handle modified dynamical equations in modified theories, very few
attempts have been made to explore axial symmetry.

The purpose of this manuscript is to work out the instability problem for
axially symmetric (in absence of reflection and radiation) anisotropic
sources in context of $f(R,T)$ gravity. The reason of avoiding reflection and
rotation terms in axially symmetry is only to somehow reduce the
complications in analysis. The modified EH action in $f(R,T)$ admits
arbitrary function of $R$ and $T$ to account the exotic matter. The action in
$f(R,T)$ is given by \cite{28}
\begin{equation}\label{1}
\int dx^4\sqrt{-g}[\frac{f(R, T)}{16\pi G}+\mathcal{L} _ {(m)}],
\end{equation}
where $\mathcal{L} _ {(m)}$ represent matter Lagrangian and $g$ represents
the metric. Several choices of $\mathcal{L} _ {(m)}$ can be considered, each
of which stands for a specific form of fluid.

The article arrangement is: The matter configuration and components of field
equations together with the dynamical equations are furnished in section
\textbf{2}. Section \textbf{3} covers the information about the $f(R,T)$
model and perturbed conservation equations leading to the collapse equation.
Section \textbf{4} contains range of stability of N and pN limits in the form
of adiabatic index. The last section consists of concluding remarks followed
by an appendix.

\section{Interior Spacetime and Dynamical Equations}

The general line element for axially symmetric compact objects constituting
five independent metric coefficients is given by
\begin{equation}\label{1}
ds^2=-A^2dt^{2}+B^2dr^{2}+B^2r^2 d\theta^{2}
+C^2d\phi^{2}+2Gdtd\theta+2Hdtd\phi,
\end{equation}
where the metric functions $A, B, C, G, H$ have dependence on time, radial
and axial coordinates $(t, r, \theta)$. Herein, we have ignored the
meridional motions and rotation about the symmetry axis. Absence of
$dtd\theta$ and $dtd\phi$ terms lead to the restricted character i.e.
vorticity free case. The modified equations are highly non-linear in nature,
so it is a tough task to handle such equations with non diagonal entries in
the metric tensor, that is why we have taken zero vorticity case to somehow
manage dynamical analysis by analytic approach.

The reduced form of general axial symmetry with three independent metric functions is \cite{32}
\begin{equation}\label{1''}
ds^2=-A^2(t,r,\theta)dt^{2}+B^2(t,r,\theta)(dr^{2}+r^2
d\theta^{2}) +C^2(t,r,\theta)d\phi^{2}.
\end{equation}
Taking $\mathcal{L} _ {(m)}=- \rho$, $8\pi G = 1$ and varying the EH action
(\ref{28}) with respect to the metric tensor $g_{uv}$ leads to the following
form for modified field equations
\begin{eqnarray}\nonumber
G_{uv}&=&\frac{1}{f_R}\left[(f_T+1)T^{(m)}_{uv}+\rho g_{uv}f_T+
\frac{f-Rf_R}{2}g_{uv}\right.\\\label{4}&+&\left.(\nabla_u\nabla_v-g_{uv}\Box)f_R\right],
\end{eqnarray}
where $\Box=\nabla^{u}\nabla_{v}$, $f_R\equiv df(R,T)/dR$, $f_T\equiv
df(R,T)/dT$, $\nabla_{u}$ is covariant derivative and  $T^{(m)}_{uv}$ is the
energy momentum tensor for usual matter. The matter configuration is
considered to be locally anisotropic \cite{25}, given by
\begin{eqnarray}\nonumber&&
T^{(m)}_{uv}=(\rho+p_\perp)V_{u}V_{v}-(K_uK_v-\frac{1}{3}h_{uv})(P_{zz}-P_{xx})-
(L_uL_v-\frac{1}{3}h_{uv})(P_{zz}\\\label{3}&&-P_{xx})+Pg_{uv}+2K_{(u}L_{v)}P_{xy},
\end{eqnarray}
where $\rho$ is the energy density and
\begin{equation}\nonumber
P=\frac{1}{3}(P_{xx}+P_{yy}+P_{zz}), \quad h_{uv}=g_{uv}+V_{u}V_{v},
\end{equation}
$P_{xx}, P_{yy}, P_{zz}$ and $P_{xy}$ are respective stresses causing
pressure anisotropy, provided that $P_{xy}=P_{yx}$ and $P_{xx}\neq P_{yy}\neq
P_{zz}$. $K_u$ and $L_u$ represents the four vectors in radial and axial
directions, respectively and $V_{u}$ is for four-velocity, these quantities
are linked as
\begin{equation}\label{4}
V_{u}=-A\delta^{0}_{u},\quad K_{u}=B\delta^{1}_{u}1,\quad
L_{u}=rB\delta^{2}_{u}.
\end{equation}
The components of modified (effective) Einstein tensor are
\begin{eqnarray}\nonumber
G^{00}&=&\frac{1}{A^2f_R}\rho+\frac{1}{A^2f_R}\left[\frac{f-Rf_R}{2}
-\frac{\dot f_R}{A^2}\left(\frac{\dot{2B}}{B}+\frac{\dot{C}}{C}\right)
-\frac{f_R'}{B^2}\left(\frac{1}{r}+\frac{2B'}{B}-\frac{C'}{C}\right)\right.\\\label{8}&&\left.
-\frac{f_R^\theta}{r^2B^2}\left(\frac{2B^\theta}{B}-\frac{C^\theta}{C}\right)
+\frac{f_R''}{B^2}\right],
\\\label{9}
G^{01}&=&\frac{-1}{A^2B^2f_R}\left[\frac{A'}{A}\dot{f_R}+\frac{\dot{B}}{B}f_R'-\dot{f_R}'
\right],
\\\label{10}
G^{02}&=&\frac{-1}{r^2A^2B^2f_R}\left[\frac{A^\theta}{A}\dot{f_R}+\frac{\dot{B}}{B}f_R^\theta-\dot{f_R}^\theta
\right],
\\\nonumber
G^{11}&=&\frac{1}{B^2f_R}\left[P_{xx}(f_T+1)+\rho f_T+\frac{\dot f_R}{A^2}
\left(\frac{\frac{\dot{B}}{B}-\dot{A}}{A}-\frac{\dot{C}}{C}\right)-
\frac{f-Rf_R}{2}-\frac{
f_R^{\theta\theta}}{r^2B^2}- \frac{\ddot{f_R}}{A^2}
\right.\\\label{11}&&\left.+\frac{f_R'}{B^2}\left(\frac{1}{r}-\frac{A'}{A}+\frac{B'}{B}
-\frac{C'}{C}\right)+
\frac{f_R^\theta}{r^2B^2}\left(\frac{3B^\theta}{B}-\frac{A^\theta}{A}-\frac{C^\theta}{C}\right)\right],
\\\label{12}
G^{12}&=&\frac{1}{r^2B^4f_R}\left[P_{xy}(f_T+1)+f_R'^\theta-
\frac{B^\theta}{B}f'_R-\frac{B'}{B}f_R^\theta\right],
\\\nonumber
G^{22}&=&\frac{1}{r^2B^4f_R}\left[P_{yy}(f_T+1)+\rho f_T+\frac{\dot f_R}{A^2}\left(\frac{\dot{B}}{B}-\frac{\dot{A}}{A}
+\frac{\dot{C}}{C}\right)+
\frac{\ddot{f_R}}{A^2}-\frac{f-Rf_R}{2}\right.\\\label{13}&&\left.-\frac{f_R''}{B^2}
-\frac{f_R^\theta}{r^2B^2}\left(\frac{A^\theta}{A}-\frac{B^\theta}{B}
+\frac{C^\theta}{C}\right)-\frac{f_R'}{B^2}\left(\frac{A'}{A}-\frac{B'}{B}+\frac{C'}{C}\right)
\right], \\\nonumber
G^{33}&=&\frac{1}{C^2f_R}\left[P_{zz}(f_T+1)+
\frac{\ddot{f_R}}{A^2}-\frac{ f_R^{\theta\theta}}{r^2B^2}+\rho f_T-\frac{f-Rf_R}{2}-\frac{\dot
f_R}{A^2} \left(\frac{\dot{A}}{A}
-\frac{\dot{2B}}{B}\right)\right.\\\label{14}&&\left.
-\frac{f_R''}{B^2}-\frac{f_R'}{B^2}\left(\frac{A'}{A}-\frac{2B'}{B}-\frac{1}{r}\right)
-\frac{f_R^\theta}{r^2B^2}\left(\frac{A^\theta}{A}-\frac{2B^\theta}{B}\right)\right].
\end{eqnarray}
Here dot, prime and $\theta$ indicate the derivatives w.r.t $t, r$ and $\theta$
coordinates respectively. The expression for Ricci scalar is
\begin{eqnarray}\nonumber
R&=&\frac{2}{A^2}\left[\frac{\dot{A}}{A}\left(\frac{\dot{2B}}{B}+\frac{\dot{C}}{C}\right)-
\frac{\dot{B}}{B}\left(\frac{\dot{B}}{B}+\frac{\dot{2C}}{C}\right)-\frac{2\ddot{B}}{B}
-\frac{\ddot{C}}{C}\right]\\\nonumber&&+\frac{2}{B^2}\left[\frac{A''}{A}+
\frac{A'C'}{AC}+\frac{B''}{B}-\frac{1}{r}\left(\frac{A'}{A}-\frac{B'}{B}-\frac{C'}{C}\right)-
\frac{B'^2}{B^2}\right.\\\label{14'}&&\left.+\frac{C''}{C}+\frac{1}{r^2}
\left(\frac{A^{\theta\theta}}{A}+
\frac{B^{\theta\theta}}{B}+\frac{C^{\theta\theta}}{C}
-(\frac{{B^\theta}}{B})^2+\frac{A^\theta
C^\theta}{AC}\right)\right].
\end{eqnarray}

In order to explore stellar evolution, one needs to arrive at dynamical
equations that can be obtained by employing contracted Bianchi identities.
Conservation laws play significant part in establishment of instability range
by more generic analytic approach, the dynamical equations in our case are
\begin{eqnarray}\label{15}
&&G^{uv}_{;v}V_{u}=0 \Rightarrow \left[\frac{1}{f_R}T^{0v}
+\frac{1}{f_R}\overset{(D)}{T^{0v}}\right]_{;v}(-A)=0,
\\\label{16}
&&G^{uv}_{;v}K_{u}=0 \Rightarrow \left[\frac{1}{f_R}T^{1v}
+\frac{1}{f_R}\overset{(D)}{T^{1v}}\right]_{;v}(B)=0,
\\\label{17}
&&G^{uv}_{;v}L_{u}=0 \Rightarrow \left[\frac{1}{f_R}T^{2v}
+\frac{1}{f_R}\overset{(D)}{T^{2v}}\right]_{;v}(rB)=0,
\end{eqnarray}
on simplification, we have
\begin{eqnarray}\nonumber &&
G^{00}_{,0}+ G^{01}_{,1}+G^{02}_{,2}+G^{00}\left(\frac{2\dot{A}}{A}
+\frac{\dot{2B}}{B}+\frac{\dot{C}}{C}\right)+G^{01}\left(\frac{3A'}{A}+\frac{2B'}{B}
+\frac{C'}{C}+\frac{1}{r}\right)\\\label{18}&&+G^{02}\left(\frac{3A^\theta}{A}+\frac{2B^\theta}{B}
+\frac{C^\theta}{C}\right)+G^{11}\frac{B\dot{B}}{A^2}+G^{22}\frac{r^2B\dot{B}}{A^2}
+G^{33}\frac{C\dot{C}}{A^2}=0,
\\\nonumber &&
G^{01}_{,0}+
G^{11}_{,1}+G^{12}_{,2}+G^{00}\frac{AA'}{B^2}+G^{01}\left(\frac{\dot{A}}{A}
+\frac{\dot{C}}{C}+\frac{\dot{4B}}{B}\right)+G^{11}\left(\frac{A'}{A}+\frac{3B'}{B}
+\frac{C'}{C}\right.\\\label{19}&&\left.+\frac{1}{r}\right)+G^{12}\left(\frac{A^\theta}{A}
+\frac{4B^\theta}{B}
+\frac{C^\theta}{C}\right)-G^{22}\left(r+\frac{r^2B'}{B}\right)+G^{33}\frac{CC'}{B^2}=0,
\end{eqnarray}
\begin{eqnarray}\nonumber &&
G^{02}_{,0}+ G^{12}_{,1}+G^{22}_{,2}+G^{00}\frac{AA^\theta}{r^2B^2}+G^{02}\left(\frac{\dot{A}}{A}
+\frac{\dot{4B}}{B}+\frac{\dot{C}}{C}\right)-\frac{B^\theta}{r^2B}G^{11}+\left(\frac{A'}{A}
\right.\\\label{20}&&\left.+\frac{4B'}{B}
+\frac{C'}{C}+\frac{3}{r}\right)G^{12}+G^{22}\left(\frac{A^\theta}{A}+\frac{3B^\theta}{B}
+\frac{C^\theta}{C}\right)-G^{33}\frac{CC^\theta}{r^2B^2}=0.
\end{eqnarray}
The notations of $0, 1$ and $2$ indicates $t, r$ and $\theta$. Terms
belonging to matter or effective part of the dynamical equations can be
viewed separately by inserting components of Einstein tensor given in Eqs.
(\ref{8})-(\ref{14}). The dynamics of gravitating axial system can be
explored with the help of perturbation scheme, that is useful in estimating
the change in system with the passage of time.

\section{$f(R, T)$ Model and Perturbation Approach}

Selection of the model under observation is a crucial constituent of the
analysis. Since we are dealing the system analytically, thus model selected
shall bring fruitful mechanism for some particular form of $f(R, T)$. We
found that the $f(R, T)$ form suitable for dynamical analysis is constrained
to $f(R,T)=f(R)+\lambda T$, where $\lambda$ is positive constant and $f(R)$
is an arbitrary function of Ricci scalar. The origin of such restriction
comes from the fact that non-linear terms of trace in $f(R, T)$ complicates
the formation of modified field equations that can not be handled
analytically. Such $f(R, T)$ models bearing non-linear terms of trace of
energy momentum can be dealt by using numerical techniques leading to more
specified outcomes, whereas, the findings of analytic approach yield more
generic. The viable $f(R, T)$ model we have chosen is
\begin{equation}\label{21}
f(R,T)=R+\alpha R^2+\lambda T,
\end{equation}
where any positive value can be assigned to $\alpha$ and $\lambda$.

The onset of modified field equations is non-linear in nature whose solution
is still undetermined that is why perturbation approach is utilized to
monitor variations in gravitating system with the time transition. All
physical quantities are taken to be time independent initially, passage of
time implicates dependence on time as well. To introduce first order
perturbations, we chose $0<\epsilon\ll1$
\begin{eqnarray}\label{22}
A(t,r,\theta)&=&A_0(r,\theta)+\epsilon D(t)a(r,\theta),\\\label{23}
B(t,r,\theta)&=&B_0(r,\theta)+\epsilon D(t)b(r,\theta),\\\label{24}
C(t,r,\theta)&=&C_0(r,\theta)+\epsilon D(t)c(r,\theta),\\\label{25}
\rho(t,r,\theta)&=&\rho_0(r,\theta)+\epsilon
{\bar{\rho}(t,r,\theta)},
\\
\label{26} P_{xx}(t,r,\theta)&=&P_{xx0}(r,\theta)+\epsilon
{\bar{P}_{xx}(t,r,\theta)},
\\\label{27}
P_{yy}(t,r,\theta)&=&P_{yy0}(r,\theta)+\epsilon {\bar{P}_{yy}(t,r,\theta)},
\\\label{28}
P_{zz}(t,r,\theta)&=&P_{zz0}(r,\theta)+\epsilon {\bar{P}_{zz}(t,r,\theta)},
\\\label{29}
P_{xy}(t,r,\theta)&=&P_{xy0}(r,\theta)+\epsilon {\bar{P}_{xy}(t,r,\theta)},
\\\label{30}
R(t,r,\theta)&=&R_0(r,\theta)+\epsilon D(t)e(r,\theta),\\\nonumber
f(R, T)&=&[R_0(r,\theta)+\alpha R_0^2(r,\theta)+\lambda T_0(r,\theta)]+\epsilon D(t)e(r,\theta)[1\\\label{51'}&+&2\alpha  R_0(r,\theta)],\\\label{52'}
f_R&=&1+2\alpha R_0(r,\theta)+\epsilon 2\alpha D(t)e(r,\theta),\\\label{52''}
f_T&=&\lambda.
\end{eqnarray}
The first order perturbed Bianchi identities (\ref{18})-(\ref{20})
imply
\begin{eqnarray}\nonumber&&
\left[\dot{\bar{\rho}} +\left\{\rho_0\left(\frac{a}{A_0}+\frac{2\lambda_1b}{B_0}+\frac{\lambda_1c}{C_0}\right)+
\frac{\lambda_1b}{B_0}(P_{xx0}+P_{yy0})+\frac{\lambda_1c}{C_0}P_{zz0}+Z_{1p}\right\}\dot{D}\right]=0,
\\\label{33}&&
\\\nonumber&&
\left[\lambda_1\bar{P_{xx}}+\lambda\bar{\rho}-2(\lambda_1{P_{xx0}}+\lambda\rho_0)\left(\frac{b}{B_0}
+\frac{e\alpha}{I}\right)D\right]_{,1}+\left(\lambda_1\bar{P_{xx}}+\lambda\bar{\rho}\right)
\left(\frac{A_0'}{A_0}+\frac{3B_0'}{B_0}
\right.\\\nonumber&&\left.+\frac{C_0'}{C_0}
+\frac{1}{r}\right)+\frac{1}{r^2}\left[\lambda_1\bar{P_{xy}}-2\left(\frac{2b}{B_0}
+\frac{e\alpha}{I}\right){P_{xy0}}D\right]_{,2}+\frac{\lambda_1\bar{P_{xy}}}{r^2B_0^2}\left(\frac{A_0^\theta}{A_0}
+4\frac{B_0^\theta}{B_0}+\frac{C_0^\theta}{C_0}\right)+\\\nonumber&&\left(\lambda_1\bar{P_{yy}}+\lambda\bar{\rho}\right)
\left(\frac{1}{r}+\frac{B_0'}{B_0}\right)+\left(\lambda_1\bar{P_{zz}}+\lambda\bar{\rho}\right)\frac{C_0'}{C_0}+
D\left[(\lambda_1{P_{xx0}}+\lambda\rho_0)\left(\left(\frac{a}{A_0}\right)'+
\left(\frac{c}{C_0}\right)'\right.\right. \\\nonumber&&\left.\left.+3\left(\frac{b}{B_0}\right)'-\left(\frac{2b}{B_0}+\frac{e\alpha}{I}\right)
\left(\frac{A_0'}{A_0}+\frac{3B_0'}{B_0}+\frac{C_0'}{C_0}+\frac{1}{r}\right)\right)+(\lambda_1{P_{yy0}}
+\lambda\rho_0)
\left(\left(\frac{b}{B_0}\right)'\right. \right. \\\nonumber&&\left.\left.-\left(\frac{2b}{B_0}+\frac{e\alpha}{I}\right)\frac{B_0'}{B_0}
\right)\left(\frac{1}{r}+\frac{B_0'}{B_0}\right)+(\lambda_1{P_{zz0}}+\lambda\rho_0)
\left(\left(\frac{c}{C_0}\right)'-\left(\frac{2b}{B_0}+\frac{e\alpha}{I}\right)\frac{C_0'}{C_0}
\right)\right. \\\label{34}&&\left.+\lambda_1{P_{xy0}}\left(\left(\frac{a}{A_0}\right)^\theta+
4\left(\frac{b}{B_0}\right)^\theta+\left(\frac{c}{C_0}\right)^\theta-\left(\frac{2b}{B_0}
+\frac{e\alpha}{I}\right)\frac{C_0^\theta}{C_0}\right)\right]+Z_{2p}=0,
\end{eqnarray}\begin{eqnarray}\nonumber&&
\left[\lambda_1\bar{P_{yy}}+\lambda\bar{\rho}-2(\lambda_1{P_{yy0}}+\lambda\rho_0)\left(\frac{b}{B_0}
+\frac{e\alpha}{I}\right)D\right]_{,2}+\left[\frac{1}{r^2B_0^4I}\lambda_1\bar{P_{xy}}\right]'
+\bar{\rho}\frac{A_0^\theta}{A_0}\\\nonumber&&
+\left(\lambda_1\bar{P_{xx}}+\lambda\bar{\rho}\right)\frac{B_0^\theta}{B_0}+
\lambda_1\bar{P_{xy}}
\left(\frac{A_0'}{A_0}+\frac{4B_0'}{B_0}+\frac{C_0'}{C_0}
+\frac{3}{r}\right)+\left(\lambda_1\bar{P_{yy}}+\lambda\bar{\rho}\right)\left(\frac{A_0^\theta}{A_0}
\right.\\\nonumber&&\left.
+3\frac{B_0^\theta}{B_0}+\frac{C_0^\theta}{C_0}\right)+\left(\lambda_1\bar{P_{zz}}
+\lambda\bar{\rho}\right)\frac{C_0^\theta}{C_0}
+
D\left[\rho_0\left(\left(\frac{a}{A_0}\right)^\theta-2\left(\frac{b}{B_0}
+\frac{e\alpha}{I}\right)\frac{A_0^\theta}{A_0}\right)\right.\\\nonumber&&\left.+
\lambda_1{P_{xy0}}\left(\left(\frac{a}{A_0}\right)'+
\left(\frac{c}{C_0}\right)'+4\left(\frac{b}{B_0}\right)'-\left(\frac{4b}{B_0}+\frac{2e\alpha}{I}\right)
\left(\frac{A_0'}{A_0}+\frac{4B_0'}{B_0}+\frac{C_0'}{C_0}
\right.\right.\right.\\\nonumber&&\left.\left.\left.+\frac{3}{r}\right)\right)+(\lambda_1{P_{xx0}}
+\lambda\rho_0)\left(
4\left(\frac{b}{B_0}\right)^\theta-\left(\frac{2b}{B_0}
+\frac{e\alpha}{I}\right)\frac{B_0^\theta}{B_0}\right)+(\lambda_1{P_{yy0}}
+\lambda\rho_0)\right.\\\nonumber&&\left.\times
\left(\left(\frac{a}{A_0}\right)^\theta+
3\left(\frac{b}{B_0}\right)^\theta+\left(\frac{c}{C_0}\right)^\theta-\left(\frac{2b}{B_0}
+\frac{e\alpha}{I}\right)\left(\frac{A_0^\theta}{A_0}+3\frac{B_0^\theta}{B_0}
+\frac{C_0^\theta}{C_0}\right)\right)
\right.\\\label{35}&&\left.+(\lambda_1{P_{zz0}}
+\lambda\rho_0)\left(\left(\frac{c}{C_0}\right)^\theta-\left(\frac{2b}{B_0}
+\frac{e\alpha}{I}\right)\frac{C_0^\theta}{C_0}\right)\right]+Z_{3p}=0,
\end{eqnarray}
where $Z_{1p}, Z_{2p}$ and $Z_{3p}$ given in appendix. To make simplification
a bit easier we substitute, $I = 1+2\alpha R_0$ and $J = e2\alpha R_0$. The
expression for energy density $\bar{\rho}$ is derived from Eq.(\ref{33}) as
\begin{eqnarray}\nonumber&&
\bar{\rho}=-\left\{\rho_0\left(\frac{a}{A_0}+\frac{2\lambda_1b}{B_0}+\frac{\lambda_1c}{C_0}\right)+
\frac{\lambda_1b}{B_0}(P_{xx0}+P_{yy0})+\frac{\lambda_1c}{C_0}P_{zz0}+Z_{1p}\right\}D.\\\label{36}&&
\end{eqnarray}

The energy density and pressure
stresses are associated as \cite{25, 33}
\begin{equation}\label{39}
\bar{P}_i=\Gamma\frac{p_{i0}}{\rho_0+p_{i0}}\bar{\rho}.
\end{equation}
where $\Gamma$ describe the variation of different stresses with energy
density. The index has variation as $i=xx, yy, xy, zz$, and Eq.(\ref{36})
together with Eq (\ref{39}) leads to corresponding perturbed stresses.
Implementation of linear perturbation on Ricci scalar yields an ordinary
differential equation having solution of following form
\begin{equation}\label{38}
D(t)=-e^{\sqrt{Z_4}t}.
\end{equation}
The expression for $Z_4$ is provided in appendix, Eq.(\ref{38}) is valid for
overall positive values of $Z_4$.

\section{N and pN Approximation}

This section constitutes the terms belonging to N and pN limits with
instability criterion in the form of adiabatic index. Making use of Eqs.
(\ref{38}) and (\ref{39}) in Eq. (\ref{34}) leads to the evolution equation.
The N and pN approximations for considered system are discussed in following
subsections.

\subsection{Newtonian Approximation}

To approximate instability/stability range in Newtonian regime, we let $A_0=1,~B_0=1$,
$\rho_0\gg p_{i0}; i=xx, yy, xy, yy$ and Schwarzschild coordinates
$C_0=r$, evolution equation along with these assumptions yield
\begin{equation}\label{n}
\Gamma <
\frac{\lambda N_0'-\frac{3}{r}N_0-2\lambda(\rho_0N_2)'-\frac{2}{r}(P_{xy0}N_2)^\theta+\lambda N_2 N_3-\frac{2}{r}N_2+\lambda
P_{xy0}N_4+Z_{2^{N}_{p}}}{\lambda_1(P_{xx0}N_1)'+\frac{\lambda_1}{r^2}
(P_{xy0}N_1)^\theta-\frac{1}{r}N_1(P_{xx0}+P_{yy0}+P_{zz0})},
\end{equation}
where $Z_{2^{N}_{p}}$ corresponds to the N-approximation terms of $Z_{2p}$,
and
\begin{eqnarray}\nonumber &&
N_0=-\left\{\rho_0N_1+
\lambda_1b(P_{xx0}+P_{yy0})+\frac{\lambda_1c}{r}P_{zz0}+Z_{1^{N}_{p}}\right\},
\\\nonumber&&
N_1=a+2\lambda_1b+\frac{\lambda_1c}{r}, \quad N_2=b+\frac{\alpha e}{I},
\\\nonumber&&
N_3=a'+4b'+2(\frac{c}{r})', \quad N_4=a^\theta+4b^\theta+\frac{c^\theta}{r}.
\end{eqnarray}
The inequality for $\Gamma$ contains both material functions and effective
part entries, system remains stable as long as the inequality (\ref{n})
holds. The terms appearing in expression for $\Gamma$ are presumed in a way
that all terms maintain positivity, this requirement impose some restrictions
on physical parameters. The constraints in N-approximation are
\begin{eqnarray}\nonumber &&
P_{xx0}+P_{yy0}+P_{zz0}<\frac{\lambda_1r}{N_1}((P_{xx0}N_1)'+\frac{1}{r^2}, \quad (P_{xy0}N_2)^\theta<-2\lambda
(\rho_0N_2)',
\end{eqnarray}
Violation of these constraints imply instability in the sources and thus lead
to gravitational collapse.

\subsection*{Post Newtonian Approximation}

In pN approximation, we assume $A_0=1-\frac{m_0}{r}$ and
$B_0=1+\frac{m_0}{r}$, corresponding inequality for range of stability is
\begin{equation}\label{pn}
\Gamma <
\frac{\lambda N_{10}'+N_9N_{10}-2\lambda(\rho_0N_6)'-\frac{2}{r}(P_{xy0}N_6)^\theta+\lambda \rho_0 N_7-\frac{3}{r}N_6+\lambda
P_{xy0}N_8+Z_{2^{pN}_{p}}}{\lambda_1(P_{xx0}N_5)'+\frac{\lambda_1}{r^2}
(P_{xy0}N_5)^\theta-\frac{1}{r}N_5(P_{xx0}+P_{yy0}+P_{zz0})+N_11},
\end{equation}
where
\begin{eqnarray}\nonumber &&
N_5=\left(\frac{ar}{r-m_0}+\frac{2\lambda_1br}{r+m_0}+\frac{\lambda_1c}{r}\right), \quad N_6=\frac{2\lambda_1br}{r+m_0}+\frac{e\alpha}{I},
\\\nonumber&&
N_7=\left(\frac{ar}{r-m_0}\right)'+4\left(\frac{br}{r+m_0}\right)'
+\left(\frac{2c}{r}\right)'-N_6\left(\frac{2}{r}+\left(\frac{m_0}{r}\right)'\frac{3r}{r+m_0}\right),
\\\nonumber&&
N_8=\left(\frac{ar}{r-m_0}\right)^\theta+\left(\frac{br}{r+m_0}\right)^\theta+\left(\frac{c}{r}\right)^\theta,
\quad
N_9=\left(\frac{3}{r}+\left(\frac{m_0}{r}\right)'\frac{3r}{r+m_0}\right),
\\\nonumber&&
N_{10}=-\left\{\rho_0N_5+
\frac{2\lambda_1br}{r+m_0}(P_{xx0}+P_{yy0})+\frac{\lambda_1c}{r}P_{zz0}+Z_{1^{pN}_{p}}\right\},
\\\nonumber&&
N_{11}=\frac{P_{xy0}N_5}{(r+m_0)^2}\left(\left(\frac{ar}{r-m_0}\right)^\theta
+\left(\frac{4br}{r+m_0}\right)^\theta\right).
\end{eqnarray}
Likewise Newtonian limit metric coefficients and effective part terms can be
constrained to maintain stability of self gravitating system. System is
stable unless above mentioned inequality holds, system collapses when
ordering relation (\ref{pn}) breaksdown. One can deduce results of GR
approximations by choosing vanishing values of $\lambda$ and $ \alpha$.

\section{Summary and Discussion}

Observational Signatures supports the argument that gravitating sources might
deviate from spherical symmetry incidently. Thus non-spherical symmetries
facilitate in examining the realistic situations such as large scale
structures, weak lensing, CMB etc. Motivating from the significance of
non-spherical symmetries, we intend to explore impact of axially symmetric
gravitating source in context of $f(R, T)$ gravity. More particularly we are
dealing with restricted axial symmetry by ignoring meridional motions and
rotation about symmetry axis. The consequence of restricted character of
spacetime leads to vorticity-free case, because absence of $dtd\theta$ and
$dtd\phi$ terms indicates that vorticity of gravitating source vanishes for
at rest observer. The metric under consideration is axially symmetric with
three independent metric functions.

Implications of axial symmetry on gravitating system has been studied
extensively in GR and modified theories of gravity. The alternative gravity
theory we have chosen to establish instability range is $f(R, T)$ gravity,
because dynamical instability of axially symmetric sources in $f(R, T)$
framework has not been ascertained yet. The model under study $f(R,
T)=R+\alpha R^2+\lambda T$ is viable for positive values of $\alpha$ and
$\lambda$. The modified field equations are obtained by varying action
(\ref{28}) for anisotropic matter distribution. The components of field
equations (\ref{8})-(\ref{14}) are used to arrive at conservation equations
(\ref{18})-(\ref{20}). These equations are of fundamental importance in
establishment of instability range analytically.

The field equations are non-linear in nature, its a difficult task to
evaluate their general solution. To count with this issue, we consider linear
perturbation of usual matter and dark source terms. The perturbed physical
quantities such as energy density and anisotropic pressure stresses are
extracted from linearly perturbed components of field equations, that are
further inserted in perturbed Bianchi identities to arrive at collapse
equation carrying both material and dark source ingredients. An ordinary
differential equation is formed from perturbed Ricci scalar, whose solution
together with evolution equation provides adiabatic index.

Adiabatic index defines range of instability for N and pN approximations
inducing some constraints on physical quantities that are provided in
previous section. Corrections to GR and $f(R)$ gravity can be determined by
setting $\alpha\rightarrow0, \lambda\rightarrow0$ and $\lambda\rightarrow0$
respectively.

\section*{Appendix}

Following equations contain linearly perturbed terms of conservation
equations and Ricci scalar respectively.
\begin{eqnarray}\setcounter{equation}{1}\nonumber&&
Z_{1p}=\frac{e}{2}-A_0^2\left\{\frac{1}{A_0^2B_0^2I^2}\left(
(2\alpha e R_0)'(1-\frac{b}{B_0})-2\alpha
eR_0\frac{A_0'}{A_0}\right)\right\}_{,1}-\frac{A_0^2}{r^2}\left\{\frac{2}{A_0^2B_0^2I^2}\right.\\\nonumber&&\left.\times\left(
(\alpha e R_0)^\theta(1-\frac{b}{B_0})-(\alpha e R_0)\frac{A_0^\theta}{A_0}\right)\right\}_{,2}
+\frac{\alpha^2R_0^3}{I}+\frac{1}{B_0^2}
\left[\frac{(e^\theta (2\alpha e R_0))^\theta}{r^2}-4\alpha\right.\\\nonumber&&\left.\times\left((R_0R_0')'+\frac{(R_0R_0^\theta)^\theta}{r^2}\right)
\left(\frac{a}
{A_0}+\frac{b}{B_0}+\frac{\alpha e R_0}{I}\right)
+I'\left\{\left(\frac{c}{C_0}\right)'-2\left(\frac{b}{B_0}\right)'\right.\right.\\\nonumber&&\left.\left.
-\frac{b}{B_0}\left(\frac{2A_0'}{A_0}+\frac{2B_0'}{B_0}-\frac{3}{r}\right)
-\frac{c}{C_0}\left(\frac{A_0'}{A_0}-\frac{C_0'}{C_0}-\frac{1}{r}\right)+
\frac{(2\alpha e R_0)}{I}\left(\frac{C_0'}{C_0}+\frac{2B_0'}{B_0}\right.\right.\right.\\\nonumber&&\left.\left.\left.
-\frac{3}{r}\right)\right\}
+(e'(2\alpha e R_0))'
+\frac{I^\theta}{r^2}\left\{\left(\frac{c}{C_0}\right)^\theta
-2\left(\frac{b}{B_0}\right)^\theta-\frac{b}{B_0}\left(\frac{2A_0^\theta}{A_0}+\frac{2B_0^\theta}{B_0}\right)
\right.\right.\\\nonumber&&\left.\left.
-\frac{c}{C_0}\left(\frac{A_0^\theta}{A_0}-\frac{C_0^\theta}{C_0}\right)+
\frac{(2\alpha e R_0)}{I}\left(\frac{C_0^\theta}{C_0}+\frac{2B_0^\theta}{B_0}\right)\right\}+(2\alpha e
R_0)'\left(\frac{C_0'}{C_0}-\frac{2B_0'}{B_0}+\frac{1}{r}\right)
\right.\\\nonumber&&\left.
+\frac{(2\alpha e R_0)^\theta}{r^2}\left(\frac{C_0^\theta}{C_0}
-\frac{2B_0^\theta}{B_0}\right)+\left(\frac{2a}{A_0}+\frac{b}{B_0}\right)\left(I''
+\frac{I^{\theta\theta}}{r^2}\right)
-\left(\frac{3A_0'}{A_0}+\frac{2B_0'}{B_0}+\frac{1}{r}\right.\right.\\\nonumber&&\left.\left.
+\frac{C_0'}{C_0}
\right)\left(\frac{(2\alpha e R_0)'}{I}(1-\frac{b}{B_0})
-\frac{A_0'}{A_0}\frac{(2\alpha e R_0)}{I}\right)
+\left(\frac{A_0^\theta}{A_0}
\frac{(2\alpha e R_0)}{I}-\frac{(2\alpha e
R_0)^\theta}{I}(1\right.\right.\\\label{a1}&&\left.\left.-\frac{b}{B_0})\right)\left(\frac{3A_0^\theta}{A_0}+\frac{C_0^\theta}{C_0}
+\frac{2B_0^\theta}{B_0}\right)\right],
\end{eqnarray}
\begin{eqnarray}\nonumber&&
Z_{2p}=\left[\left[\frac{1}{IB_0^2}\left\{\frac{\ddot{D}}{DA^2_0}-\frac{1}{B_0^2}
\left\{\frac{eB_0^2}{2}+I'\left(\left(\frac{a}{A_0}\right)'-\left(\frac{b}{B_0}\right)'
+\left(\frac{c}{C_0}\right)'\right)
\right.\right.\right.\right.\\\nonumber&&\left.\left.\left.\left.
+\left(J'-\frac{2b}{B_0}I'\right)\left(\frac{A_0'}{A_0}+\frac{C_0'}{C_0}
-\frac{B_0'}{B_0}-\frac{1}{r}\right)+\frac{1}{r^2}\left(J^{\theta\theta}+\left(J^\theta
-\frac{2b}{B_0}I^\theta\right)
\left(\frac{A_0^\theta}{A_0}
\right.\right.\right.\right.\right.\right.\\\nonumber&&\left.\left.\left.\left.\left.\left.-
\frac{3B_0^\theta}{B_0}
+\frac{C_0^\theta}{C_0}\right)+
\frac{2b}{B_0}I^{\theta\theta}+I^\theta
\left(\left(\frac{a}{A_0}\right)^\theta+\left(\frac{c}{C_0}\right)^\theta-3\left(\frac{b}{B_0}\right)^\theta
\right)\right)\right\}\right\}\right]_{,1}\right.\\\nonumber&&\left.+
\left[\frac{1}{r^2IB_0^4}\left\{J'^{\theta}+\left(\frac{b}{B_0}\right)^\theta I'+
J^{\theta}\left(\frac{B_0'}{B_0}+\frac{1}{r}\right)
-\left(\frac{b}{B_0}\right)'I^\theta\right\}\right]_{,2}\right]IB_0^4
\\\nonumber&&-e\frac{B_0'}{B_0}+\frac{A_0'}{A_0}\left[J''+\frac{J^{\theta\theta}}{r^2}
-\frac{2b}{B_0}\left(I''+\frac{I^{\theta\theta}}{r^2}\right)+\left(J'
-\frac{2b}{B_0}I'\right)\left(\frac{C_0'}{C_0}
-\frac{2B_0'}{B_0}\right.\right.\\\nonumber&&\left.\left.
+\frac{1}{r}\right)+I'\left(\left(\frac{c}{C_0}\right)'-\left(\frac{b}{B_0}\right)'
\right)+\frac{1}{r^2}\left\{I^\theta
\left(\left(\frac{c}{C_0}\right)^\theta-2\left(\frac{b}{B_0}\right)^\theta
\right)+\left(J^\theta\right.\right.\right.\\\nonumber&&\left.\left.\left.
-\frac{2b}{B_0}I^\theta\right)\left(\frac{C_0^\theta}{C_0}-\frac{2B_0^\theta}{B_0}\right)\right\}\right]+
\left(\frac{(aA_0)'}{A_0^2}-\frac{2b}{B_0}\frac{A_0'}{A_0}\right)\left(\frac{\alpha R_0^2B_0^2}{2}+I''+
I'\left(\frac{C_0'}{C_0}\right.\right.\\\nonumber&&\left.\left.
-\frac{2B_0'}{B_0}+\frac{1}{r}\right)+\frac{I^{\theta\theta}}{r^2}
+\frac{I^{\theta}}{r^2}\left(\frac{C_0^\theta}{C_0}-\frac{2B_0^\theta}{B_0}\right)\right)-
\left\{\frac{\alpha R_0^2B_0^2}{2}+I'\left(\frac{A_0'}{A_0}+\frac{C_0'}{C_0}
-\frac{B_0'}{B_0}\right.\right.\\\nonumber&&\left.\left.-\frac{1}{r}\right)+\frac{I^{\theta\theta}}{r^2}
+\frac{I^{\theta}}{r^2}\left(\frac{A_0^\theta}{A_0}+\frac{C_0^\theta}{C_0}-\frac{3B_0^\theta}{B_0}\right)\right\}
\left(\left(\frac{a}{A_0}\right)'+3\left(\frac{b}{B_0}\right)'
+\left(\frac{c}{C_0}\right)'\right)\\\nonumber&&-\left(\frac{A_0'}{A_0}+\frac{C_0'}{C_0}
+\frac{3B_0'}{B_0}+\frac{1}{r}\right)\left\{I'\left(\left(\frac{a}{A_0}\right)'-\left(\frac{b}{B_0}\right)'
+\left(\frac{c}{C_0}\right)'\right)
+\left(\frac{A_0'}{A_0}
-\frac{B_0'}{B_0}\right.\right.\\\nonumber&&\left.\left.
+\frac{C_0'}{C_0}-\frac{1}{r}\right)\left(J'-\frac{2b}{B_0}I'\right)+\frac{1}{r^2}\left(J^{\theta\theta}+\left(J^\theta
-\frac{2b}{B_0}I^\theta\right)
\left(\frac{A_0^\theta}{A_0}-
\frac{3B_0^\theta}{B_0}
+\frac{C_0^\theta}{C_0}\right)\right.\right.\\\nonumber&&\left.\left.+
\frac{2b}{B_0}I^{\theta\theta}+I^\theta
\left(\left(\frac{a}{A_0}\right)^\theta+\left(\frac{c}{C_0}\right)^\theta
-3\left(\frac{b}{B_0}\right)^\theta
\right)\right)\right\}
-\left[\left(\left(\frac{a}{A_0}\right)^\theta
+\left(\frac{c}{C_0}\right)^\theta
\right.\right.\\\nonumber&&\left.\left.+4\left(\frac{b}{B_0}\right)^\theta
\right)\left(I'^\theta +\frac{B_0^\theta}{B_0}I'+I^\theta\left(\frac{B_0'}{B_0}
+\frac{1}{r}\right)\right)-\left(\frac{A_0^\theta}{A_0}+
\frac{4B_0^\theta}{B_0}
+\frac{C_0^\theta}{C_0}\right)\left(\frac{B_0^\theta}{B_0}J'
\right.\right.\\\nonumber&&\left.\left.
-J'^\theta- I'\left(\frac{b}{B_0}\right)^\theta -J^\theta\left(\frac{B_0'}{B_0}
+\frac{1}{r}\right)+I^\theta\left(\frac{b}{B_0}\right)'\right)\right]\frac{1}{r^2}-
\left(\frac{B_0'}{B_0}+\frac{1}{r}\right)\left[\frac{B_0^2}{A_0^2}\frac{\ddot{D}}{D}J
\right.\\\nonumber&&\left.-J''+\frac{2b}{B_0}I''-I'\left(\left(\frac{a}{A_0}\right)'-\left(\frac{b}{B_0}\right)'
+\left(\frac{c}{C_0}\right)'\right)
-\left(J'-\frac{2b}{B_0}I'\right)\left(\frac{A_0'}{A_0}+\frac{C_0'}{C_0}\right.\right.
\end{eqnarray}
\begin{eqnarray}
\nonumber&&\left.\left.-\frac{B_0'}{B_0}\right)+\frac{1}{r^2}\left(I^\theta
\left(\left(\frac{a}{A_0}\right)^\theta+\left(\frac{c}{C_0}\right)^\theta-\left(\frac{b}{B_0}\right)^\theta
\right)-\left(J^\theta -\frac{2b}{B_0}I^\theta\right)
\left(\frac{A_0^\theta}{A_0}-\frac{B_0^\theta}{B_0}
\right.\right.\right.\\\nonumber&&\left.\left.\left.+\frac{C_0^\theta}{C_0}\right)\right)\right]
+\left(\frac{b}{B_0}\right)'\left[\frac{LB_0^2}{2}+I'
\left(\frac{A_0'}{A_0}+\frac{C_0'}{C_0}-\frac{B_0'}{B_0}\right)-\frac{I^\theta}{r^2}
\left(\frac{A_0^\theta}{A_0}-\frac{B_0^\theta}{B_0}
+\frac{C_0^\theta}{C_0}\right)\right.\\\nonumber&&\left.-I''\right]+
\frac{C_0'}{C_0}\left[J''-\frac{2b}{B_0}I''+I'\left(\left(\frac{a}{A_0}\right)'-\left(\frac{2b}{B_0}\right)'\right)
+\left(J'-\frac{2b}{B_0}I'\right)\left(\frac{A_0'}{A_0}-\frac{B_0'}{B_0}\right.\right.\\\nonumber&&\left.\left.
+\frac{1}{r}\right) +\frac{1}{r^2}\left\{J^{\theta\theta}+
\left(J^\theta -\frac{2b}{B_0}I^\theta\right)
\left(\frac{A_0^\theta}{A_0}-\frac{2B_0^\theta}{B_0}\right)+I^\theta
\left(\left(\frac{a}{A_0}\right)^\theta-2\left(\frac{b}{B_0}\right)^\theta
\right)\right.\right.\\\nonumber&&\left.\left.
-\frac{2b}{B_0}I^{\theta\theta}\right\}\right]+\left(\frac{(cC_0)'}{C_0^2}-\frac{2b}{B_0}\frac{C_0'}{C_0}\right)
\left[I'
\left(\frac{A_0'}{A_0}-\frac{2B_0'}{B_0}+\frac{1}{r}\right)-\frac{I^\theta}{r^2}
\left(\frac{A_0^\theta}{A_0}-\frac{2B_0^\theta}{B_0}\right)
\right.\\\label{a2}&&\left.+\frac{\alpha R_0^2B_0^2}{2}+I''+\frac{I^{\theta\theta}}{r^2}\right]-
\frac{\ddot{D}B_0^2}{DA^2_0I}\left(J'-\frac{A_0'}{A_0}J-\frac{b}{B_0}I'\right),
\\\nonumber&&
Z_{3p}=Ir^2B_0^4\left[\left[\frac{1}{r^2IB_0^4}\left\{J'^{\theta}+\left(\frac{b}{B_0}\right)^\theta
I'+ J^{\theta}\left(\frac{B_0'}{B_0}+\frac{1}{r}\right)
-\left(\frac{b}{B_0}\right)'I^\theta\right\}\right]_{,1}\right.\\\nonumber&&\left.+
\frac{\ddot{D}B_0^2}{DA^2_0I}\left(\frac{A_0^\theta}{A_0}J+\frac{b}{B_0}I^\theta-J^\theta\right)
+\left[\frac{1}{Ir^2B_0^4}\left\{\frac{\ddot{D}B_0^2}{DA_0^2}J
-J''+\frac{2b}{B_0}I''+\left(\frac{2b}{B_0}I'\right.\right.\right.\right.
\\\nonumber&&\left.\left.\left.\left.-J'\right)\left(\frac{A_0'}{A_0}+\frac{C_0'}{C_0}
-\frac{B_0'}{B_0}\right)-I'\left(\left(\frac{a}{A_0}\right)'-\left(\frac{b}{B_0}\right)'
+\left(\frac{c}{C_0}\right)'\right)
+\frac{1}{r^2}\left(\left(\frac{2b}{B_0}I^\theta
\right.\right.\right.\right.\right.
\\\nonumber&&\left.\left.\left.\left.\left.-J^\theta
\right)
\left(\frac{A_0^\theta}{A_0}-\frac{B_0^\theta}{B_0}+\frac{C_0^\theta}{C_0}\right)-I^\theta
\left(\left(\frac{a}{A_0}\right)^\theta+\left(\frac{c}{C_0}\right)^\theta-\left(\frac{b}{B_0}\right)^\theta
\right)\right)\right\}\right]_{,2}\right]
\\\nonumber&&-e\frac{B_0^\theta}{B_0}+\frac{A_0^\theta}{A_0}\left[J''+\frac{J^{\theta\theta}}{r^2}
-\frac{2b}{B_0}\left(I''+\frac{I^{\theta\theta}}{r^2}\right)+\left(J'
-\frac{2b}{B_0}I'\right)\left(\frac{C_0'}{C_0}
-\frac{2B_0'}{B_0}\right.\right.\\\nonumber&&\left.\left.
+\frac{1}{r}\right)+I'\left(\left(\frac{c}{C_0}\right)'-\left(\frac{b}{B_0}\right)'
\right)+\frac{1}{r^2}\left\{I^\theta
\left(\left(\frac{c}{C_0}\right)^\theta-2\left(\frac{b}{B_0}\right)^\theta
\right)+\left(J^\theta\right.\right.\right.\\\nonumber&&\left.\left.\left.
-\frac{2b}{B_0}I^\theta\right)\left(\frac{C_0^\theta}{C_0}-\frac{2B_0^\theta}{B_0}\right)\right\}\right]
+\left(\frac{(aA_0)^\theta}{A_0^2}-\frac{2b}{B_0}\frac{A_0^\theta}{A_0}\right)\left(\frac{\alpha R_0^2B_0^2}{2}+I''+
I'\left(\frac{C_0'}{C_0}\right.\right.
\end{eqnarray}
\begin{eqnarray}\nonumber&&\left.\left.
-\frac{2B_0'}{B_0}+\frac{1}{r}\right)+\frac{I^{\theta\theta}}{r^2}
+\frac{I^{\theta}}{r^2}\left(\frac{C_0^\theta}{C_0}-\frac{2B_0^\theta}{B_0}\right)\right)
-\left(\frac{b}{B_0}\right)^\theta\left\{\frac{\alpha R_0^2B_0^2}{2}+I'\left(\frac{A_0'}{A_0}+\frac{C_0'}{C_0}
\right.\right.\\\nonumber&&\left.\left.-\frac{B_0'}{B_0}+\frac{1}{r}\right)+\frac{I^{\theta\theta}}{r^2}
+\frac{I^{\theta}}{r^2}\left(\frac{A_0^\theta}{A_0}+\frac{C_0^\theta}{C_0}-\frac{3B_0^\theta}{B_0}\right)\right\}
-\frac{B_0^\theta}{B_0}\left\{I'\left(\left(\frac{a}{A_0}\right)'-\left(\frac{b}{B_0}\right)'
\right.\right.\\\nonumber&&\left.\left.
+\left(\frac{c}{C_0}\right)'\right) +\left(\frac{A_0'}{A_0}
-\frac{B_0'}{B_0}
+\frac{C_0'}{C_0}-\frac{1}{r}\right)\left(J'-\frac{2b}{B_0}I'\right)+\frac{1}{r^2}\left(J^{\theta\theta}
-\left(\frac{2b}{B_0}I^\theta-\right.\right.\right.\\\nonumber&&\left.\left.\left.J^\theta\right)
\left(\frac{A_0^\theta}{A_0}- \frac{3B_0^\theta}{B_0}
+\frac{C_0^\theta}{C_0}\right)+
\frac{2b}{B_0}I^{\theta\theta}+I^\theta
\left(\left(\frac{a}{A_0}\right)^\theta+\left(\frac{c}{C_0}\right)^\theta
-3\left(\frac{b}{B_0}\right)^\theta \right)\right)\right\}
\\\nonumber&&
-\frac{1}{r^2}\left[\left(\left(\frac{a}{A_0}\right)'
+\left(\frac{c}{C_0}\right)'+\left(\frac{b}{B_0}\right)'
\right)\left(I'^\theta +\frac{B_0^\theta}{B_0}I'+I^\theta\left(\frac{B_0'}{B_0}
+\frac{1}{r}\right)\right)-\left(\frac{A_0'}{A_0}\right.\right.\\\nonumber&&\left.\left.+
\frac{4B_0'}{B_0}
+\frac{C_0'}{C_0}\right)\left(\frac{B_0^\theta}{B_0}J'
-J'^\theta- I'\left(\frac{b}{B_0}\right)^\theta -J^\theta\left(\frac{B_0'}{B_0}
+\frac{1}{r}\right)+I^\theta\left(\frac{b}{B_0}\right)'\right)\right]
\\\nonumber&&+\left(\frac{A_0^\theta}{A_0}+
\frac{3B_0^\theta}{B_0}
+\frac{C_0^\theta}{C_0}\right)\left[\frac{B_0^2}{A_0^2}\frac{\ddot{D}}{D}J+
\frac{2b}{B_0}I''-I'\left(\left(\frac{a}{A_0}\right)'-\left(\frac{b}{B_0}\right)'
+\left(\frac{c}{C_0}\right)'
\right)\right.\\\nonumber&&\left.-J''
-\left(J'-\frac{2b}{B_0}I'\right)\left(\frac{A_0'}{A_0}+\frac{C_0'}{C_0}
-\frac{B_0'}{B_0}\right)-\frac{1}{r^2}\left(\left(J^\theta
-\frac{2b}{B_0}I^\theta\right)
\left(\frac{A_0^\theta}{A_0}-\frac{B_0^\theta}{B_0}
\right.\right.\right.
\\
\nonumber&&\left.\left.\left.+\frac{C_0^\theta}{C_0}\right)-I^\theta
\left(\left(\frac{a}{A_0}\right)^\theta+\left(\frac{c}{C_0}\right)^\theta-\left(\frac{b}{B_0}\right)^\theta
\right)\right)\right]+
\left(\left(\frac{a}{A_0}\right)^\theta+\left(\frac{c}{C_0}\right)^\theta
\right.\\\nonumber&&\left.+3\left(\frac{b}{B_0}\right)^\theta
\right)\left[\frac{\alpha R_0^2B_0^2}{2}+I'
\left(\frac{A_0'}{A_0}+\frac{C_0'}{C_0}-\frac{B_0'}{B_0}\right)-\frac{I^\theta}{r^2}
\left(\frac{A_0^\theta}{A_0}-\frac{B_0^\theta}{B_0}
+\frac{C_0^\theta}{C_0}\right)-I''\right]
\\\nonumber&&-\frac{C_0^\theta}{C_0}\left[J''-\frac{2b}{B_0}I''
+I'\left(\left(\frac{a}{A_0}\right)'-\left(\frac{2b}{B_0}\right)'\right)
+\left(J'-\frac{2b}{B_0}I'\right)\left(\frac{A_0'}{A_0}-\frac{B_0'}{B_0}\right.\right.\\\nonumber&&\left.\left.
+\frac{1}{r}\right)+\frac{1}{r^2}\left\{\left(J^\theta-\frac{2b}{B_0}I^\theta\right)
\left(\frac{A_0^\theta}{A_0}-\frac{2B_0^\theta}{B_0}\right)+I^\theta
\left(\left(\frac{a}{A_0}\right)^\theta-2\left(\frac{b}{B_0}\right)^\theta
\right)\right.\right.\\\nonumber&&\left.\left.
+J^{\theta\theta}-\frac{2b}{B_0}I^{\theta\theta}\right\}\right]+\left(\frac{(cC_0)^\theta}{C_0^2}
-\frac{2b}{B_0}\frac{C_0^\theta}{C_0}\right)
\left[\frac{\alpha R_0^2B_0^2}{2}+I'\left(\frac{A_0'}{A_0}
-\frac{2B_0'}{B_0}+\frac{1}{r}\right)
\right.\\\label{a3}&&\left.+I''+\frac{I^{\theta\theta}}{r^2}-\frac{I^\theta}{r^2}r
\left(\frac{A_0^\theta}{A_0}-\frac{2B_0^\theta}{B_0}\right)\right],
\end{eqnarray}
\begin{eqnarray}\nonumber &&
Z_4=\frac{A_0^2}{2}\left(\frac{B_0C_0}{bC_0-cB_0}\right)\left[\frac{2}{B_0^2}
\left\{\frac{A_0'C_0'}{A_0C_0}\left(\frac{a'}{A'_0}-\frac{a}{A_0}+\frac{c'}{C'_0}
-\frac{c}{C_0}\right)+\frac{A_0''}{A_0}\left(\frac{a''}{A''_0}
\right.\right.\right.\\\nonumber &&\left.\left.\left.-\frac{a}{A_0}\right)
+\frac{B_0''}{B_0}\left(\frac{b''}{B''_0}-\frac{b}{B_0}\right)+\frac{C_0''}{C_0}\left(\frac{c''}{C''_0}
-\frac{c}{C_0}\right)-
\frac{1}{r}\left(\frac{a}{A_0}-\frac{b}{B_0}-\frac{c}{C_0}\right)'
\right.\right.\\\nonumber &&\left.\left.-\frac{2B_0'}{B_0}\left(\frac{b}{B_0}\right)'+\frac{2}{r^2}\left\{
\frac{2B_0^\theta}{B_0}\left(\frac{b}{B_0}\right)^\theta+
\frac{A_0^{\theta\theta}}{A_0}\left(\frac{a^{\theta\theta}}{A^{\theta\theta}_0}-\frac{a}{A_0}\right)
+\frac{B_0^{\theta\theta}}{B_0}\left(\frac{b^{\theta\theta}}{B^{\theta\theta}_0}
-\frac{b}{B_0}\right)\right.\right.\right.\\\label{a4} &&\left.\left.\left.+\frac{C_0^{\theta\theta}}{C_0}\left(\frac{c^{\theta\theta}}{C^{\theta\theta}_0}
-\frac{c}{C_0}\right)+\frac{A_0^\theta C_0^\theta}{A_0C_0}\left(\frac{a^\theta}{A^\theta_0}-\frac{a}{A_0}+\frac{c^\theta}{C^\theta_0}
-\frac{c}{C_0}\right)\right\}\right\}-e-\frac{2bR_0}{B_0}\right].
\end{eqnarray}

\end{document}